\documentclass{jkas}

\usepackage{url}

\def\year{2016} 
\def\month{Preprint} 
\def\volume{49} 
\def\issue{4} 
\def\beginpage{1} 
\def\endpage{241} 
\def\received{June 22, 2016} 
\def\accepted{July 12, 2016} 
\date{Received \received ; accepted \accepted}

\usepackage{flushend} 

\title{
Optical properties of amorphous alumina dust in the envelopes around O-rich AGB stars
}

\author[]{Kyung-Won Suh}

\affil[]
{Department of Astronomy and Space Science, Chungbuk National University,
Cheongju 28644, Korea \email{kwsuh@chungbuk.ac.kr}}

\def\corrauthor{
K.-W. Suh
}

\def\runningauthor{
Suh
}

\def\runningtitle{
Optical properties of amorphous alumina dust around O-rich AGB stars
}

\def\keywords{
{stars: AGB and post-AGB $-$ dust: extinction $-$ circumstellar matter
$-$ infrared: stars $-$ radiative transfer }
}

\def\abstracttext{
We investigate optical properties of amorphous alumina (Al$_2$O$_3$) dust
grains in the envelopes around O-rich asymptotic giant branch (AGB) stars
considering the laboratory measured optical data. We derive the optical
constants of amorphous alumina in a wide wavelength range that satisfy the
Kramers-Kronig relation and reproduce the laboratory measured data. Using the
amorphous alumina and silicate dust, we compare the radiative transfer model
results with the observed spectral energy distributions. Comparing the
theoretical models with the observations on various IR two-color diagrams for a
large sample O-rich AGB stars, we find that the amorphous alumina dust (about
10-40 \%) mixed with amorphous silicate can reproduce much more observed points
for the O-rich AGB stars with thin dust envelopes.
}

\begin{document}
\jkashead 


\section{Introduction} \label{introduction}

The main site of dust formation is believed to be the cool envelopes around
asymptotic giant branch (AGB) stars. O-rich AGB stars (M-type Miras and OH/IR
stars) typically show conspicuous 10 $\mu$m and 18 $\mu$m features in emission
or absorption. They suggest the presence of amorphous silicate dust grains in
the outer envelopes around them (Jones \& Merrill 1976).

Low mass-loss rate O-rich AGB (LMOA) stars with thin dust envelopes show the 10
$\mu$m and 18 $\mu$m emission features of amorphous silicate. High mass-loss
rate O-rich AGB (HMOA) stars with thick dust envelopes show the absorbing
features at the same wavelengths (e.g., Suh 2004). Water-ice was found in some
HMOA stars (Justtanont et al. 2006; Suh \& Kwon 2013) and amorphous alumina
(Al$_2$O$_3$) dust grains were detected in many LMOA stars (e.g., Sloan \&
Price 1998).

In a number of previous works (e.g., Markwick-Kemper et al. 2007; Suh \& Kwon
2011), the term corundum was erroneously assigned to different kinds of solid
alumina regardless of its crystal structure. However, only one of the
crystalline Al$_2$O$_3$ polymorphs deserves the name corundum: namely
$\alpha$-Al$_2$O$_3$ which has trigonal lattice symmetry (e.g., Koike et al.
1995). While the amorphous aluminum oxide material (Al$_2$O$_3$) synthesized by
Begemann et al. (1997) shows a single peak at 11.8 $\mu$m, corundum shows much
sharper multiple peaks around 12 $\mu$m (Koike et al. 1995; Zeidler et al.
2013).

In this work, we investigate optical properties of amorphous alumina
(Al$_2$O$_3$) dust in the envelopes around O-rich AGB stars. We derive the
optical constants of the alumina dust in a wide wavelength range, which satisfy
the Kramers-Kronig relation and reproduce the laboratory measured optical data.
Using the opacity function of the amorphous alumina dust, we compare the
theoretical radiative transfer model results with the observed spectral energy
distributions (SEDs) and observations on various IR two-color diagrams (2CDs)
for a large sample O-rich AGB stars.

\section{Amorphous alumina dust}

Dust opacity is determined from the optical constants
($m_{\lambda}=n_{\lambda}+ik_{\lambda}$), shape, and size of a dust grain. The
optical constants may be expressed by the complex index of refraction,
$m=n+ik$, where $\varepsilon=m^2$. We can derive two functions of complex
dielectric constants, $\varepsilon_1 (\lambda)$ and $\varepsilon_2 (\lambda)$,
from the opacity function $Q_{ext} (\lambda)$ considering the required
supplementary physical constraint. Namely, the dielectric constants should
satisfy the Kramers-Kronig relation (e.g., Bohren \& Huffman 1983).

Silicate dust is believed to be the main dust species for O-rich AGB stars
(Jones \& Merrill 1976; Suh 1999). Alumina dust has been detected in the
spectra of many LMOA stars in our Galaxy (Sloan \& Price 1998; Speck et al.
2000) and the Large Magellanic Cloud (Jones et al. 2014). The shape of the 10
$\mu$m emission feature of LMOA stars, which is mainly produced by silicates,
can be modified by addition of the alumina dust.

The broad 10 $\mu$m emission feature in $\lambda$ = 8$-$15 $\mu$m is best
fitted with a mixture of amorphous alumina and silicate dust. Suh \& Kwon
(2011) found that the dust opacity using the alumina dust as well as silicates
can improve the model fit on IR 2CDs for a large sample of LMOA stars. Egan \&
Sloan (2001) and Jones et al. (2014), who modeled the alumina abundance of
O-rich AGB stars, found that the 10 $\mu$m absorption feature is likely to be
produced by only silicate dust for HMOA stars.

Begemann et al. (1997; hereafter Be97) presented the optical constants for
amorphous alumina (Al$_2$O$_3$) in the wavelength range 7.8$-$500 $\mu$m. The
alumina grains produce a single peak at 11.8 $\mu$m and influences the shape of
the SED at around 10 $\mu$m. The bulk density of the amorphous alumina is 3.2 g
cm$^{-3}$.

Suh \& Kwon (2011) and Jones et al. (2014) used the optical constants for
amorphous (porous) alumina grains obtained by Be97 in the wavelength range
7.8$-$500 $\mu$m, which are extended to the shorter wavelength range ($\lambda$
$<$ 7.8 $\mu$m) by concatenation with the optical constants of corundum dust
measured by Koike et al. (1995; hereafter Ko95). However, the interpolated
optical constants are not physically reliable because they are from different
materials and do not satisfy the Kramers-Kronig relation.

\begin{figure*}[!t]
\centering
\includegraphics[width=140mm]{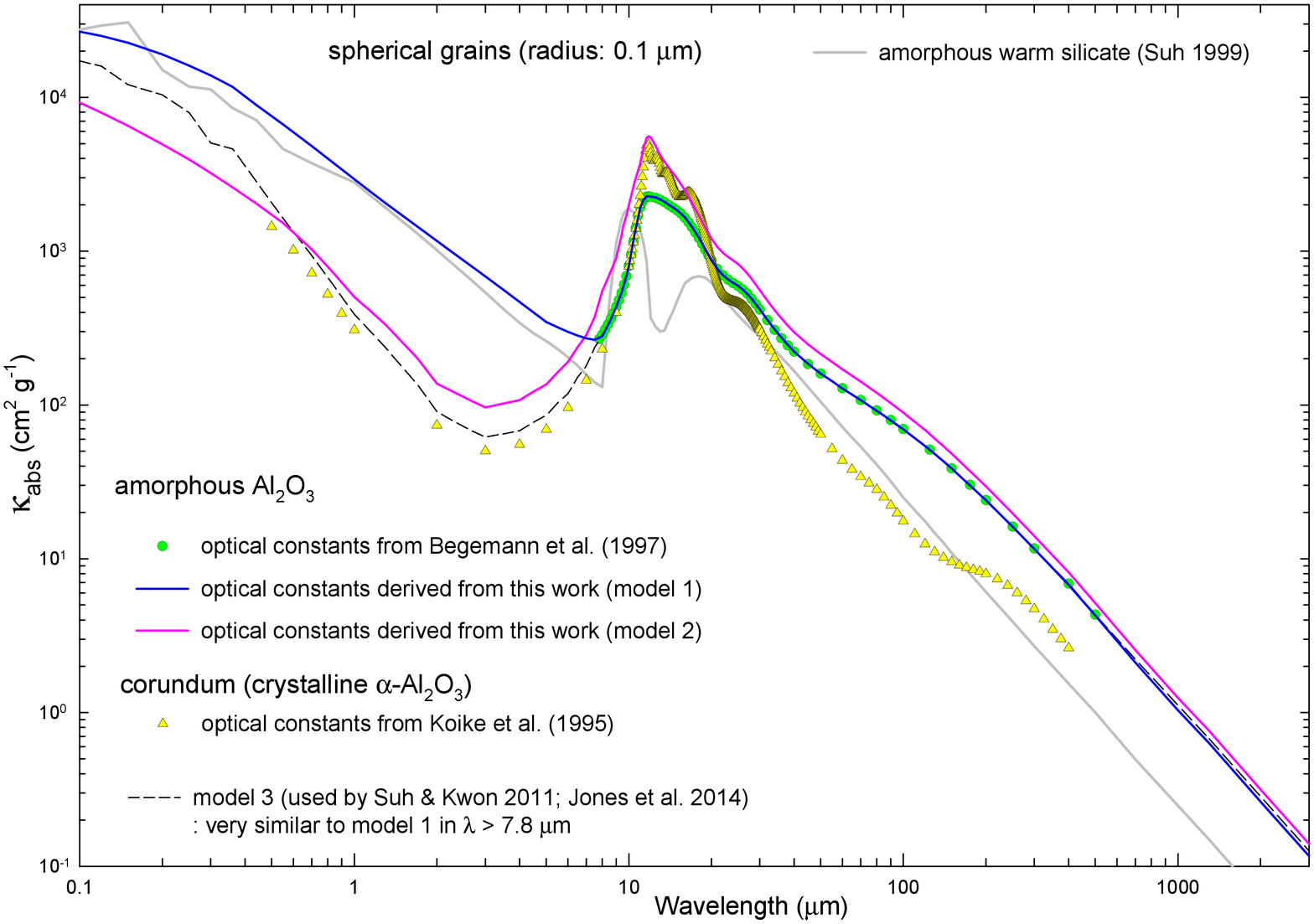} \\
\includegraphics[width=140mm]{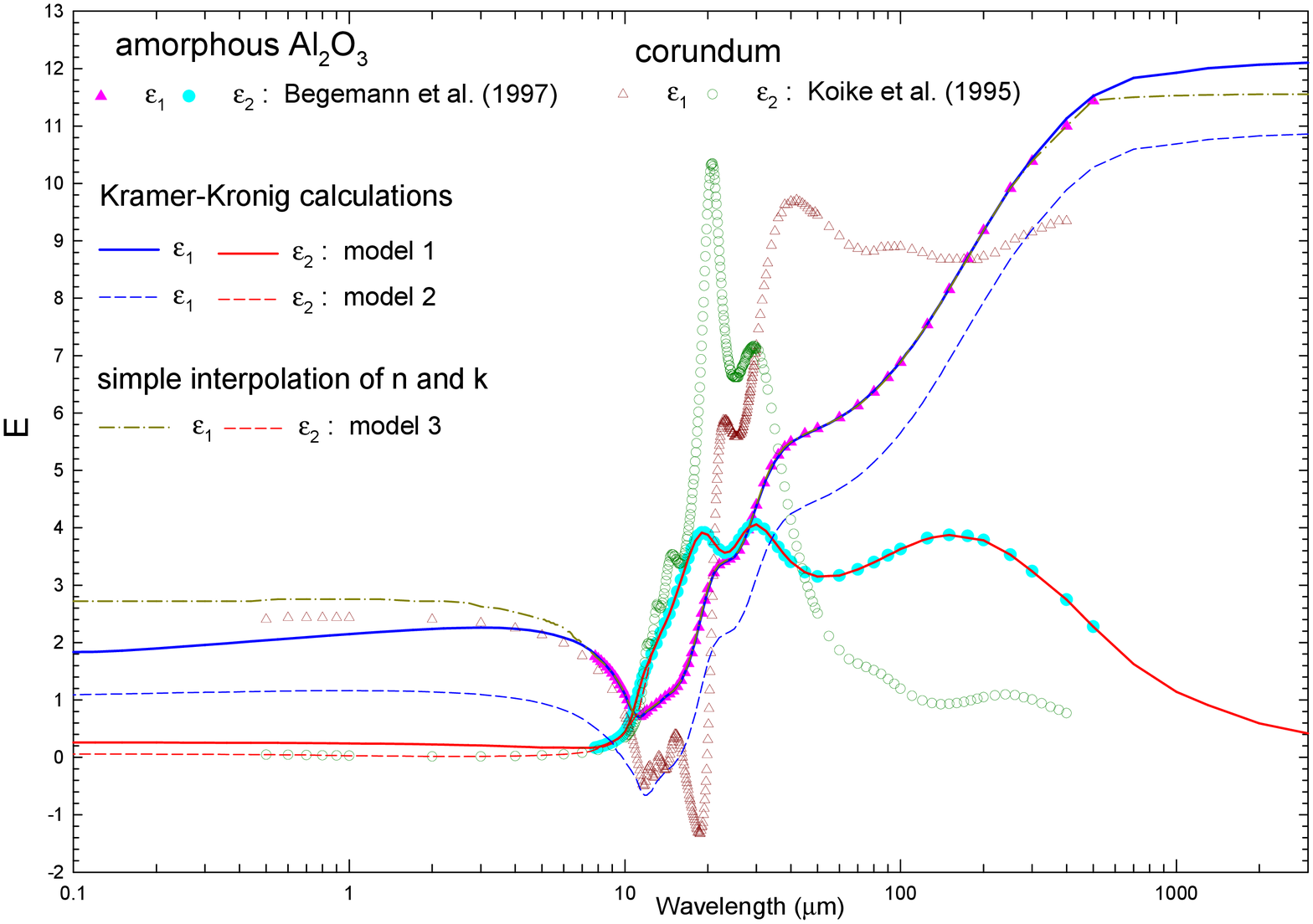}
\caption{Opacity functions and indices of refraction of for alumina dust.}
\end{figure*}

\subsection{Deriving the optical constants}

To obtain the complex dielectric functions, we have used the following
procedure similar to that used in Suh (1999):

(i) $\varepsilon_2 (\lambda)$ is adopted over a electromagnetic spectrum to
agree with the existing laboratory data.

(ii) $\varepsilon_1 (\lambda)$ is obtained using the Kramers-Kronig relation
\begin{equation}
\varepsilon _1 (\lambda) = 1 + {2 \over \pi} P \int _0^\infty
{x \varepsilon_2(x) \over x^2 - \lambda^2} dx
\end{equation}
where P indicates the Cauchy principal value.

(iii) From the complex dielectric function, we calculate $\kappa_{abs}
(\lambda)$. $\kappa_{abs} (\lambda)$ is compared with the desired opacity
function.

(iv) Where disagreements is found, the choice of $\varepsilon_2 (\lambda)$ is
modified, and steps (ii) - (iv) are repeated.

From the satisfactory sets of complex dielectric constants, we can calculate
the optical constants ($m(\lambda)=n+ik$).

\subsection{Models for amorphous alumina dust}

Figure 1 shows the opacity function (mass absorption coefficients) and complex
dielectric constants for three different models of alumina dust. The opacity
function of amorphous warm silicate (Suh 1999) is also displayed for
comparison. Using the Mie theory (Bohren \& Huffman 1983), the opacity
functions are calculated for spherical dust grains whose uniform radius is 0.1
$\mu$m. The opacity function for amorphous (porous) alumina dust measured by
Be97 shows a single peak at 11.8 $\mu$m.

For model 1, we have derived the sets of complex dielectric constants in the
wavelength range 0.1$-$3600 $\mu$m using the Kramers-Kronig procedures
described in section 2.1. We use the same $\varepsilon_2 (\lambda)$ in the
wavelength range 7.8$-$500 $\mu$m as presented by Be97 for amorphous (porous)
alumina (Al$_2$O$_3$). In other wavelength ranges, we have derived
$\varepsilon_2 (\lambda)$ so that the calculated $\varepsilon_1 (\lambda)$ in
the wavelength range 7.8$-$500 $\mu$m is almost the same as the one presented
by Be97. Small deviations of $\varepsilon_1 (\lambda)$ around the end points
(7.8 and 500 $\mu$m) were not avoidable. In the shorter (0.1$-$7.8$\mu$m)
wavelength range, where the laboratory data are not available, the opacity
function is similar to that of silicate dust. In the longer (500$-$3600 $\mu$m)
wavelength range, where the laboratory data are not available, the opacity
function obeys approximately a simple power law ($\kappa_{abs} \propto
\lambda^{-2}$) just like silicate dust.

For model 2, we use $\varepsilon_2 (\lambda)$ from Be97 (amorphous porous
alumina) in the wavelength range 7.8$-$500 $\mu$m and from Ko95 (corundum) in
the shorter wavelength range ($\lambda$ $<$ 7.8 $\mu$m). $\varepsilon_1
(\lambda)$ is calculated from the combined $\varepsilon_2 (\lambda)$ using the
Kramers-Kronig relation. Because of the small change of $\varepsilon_2
(\lambda)$ from model 1 in $\lambda$ $<$ 7.8 $\mu$m (see the lower panel of
Figure 1), the calculated $\varepsilon_1 (\lambda)$ and opacity function for
model 2 in the wavelength range 7.8$-$500 $\mu$m are very different from those
for model 1 (or Be97). Model 2 would not be physically reliable because the
complex dielectric constants are from different materials.

For model 3, the optical constants (n and k) for amorphous (porous) alumina
grains obtained by Be97 are extended to the shorter wavelength range ($\lambda$
$<$ 7.8 $\mu$m) by concatenation with the optical constants of corundum
measured by Ko95. $\varepsilon_2 (\lambda)$ is the same as the one from Ko95,
but $\varepsilon_1 (\lambda)$ shows some deviations in the range $\lambda$ $<$
7.8 $\mu$m (see Figure 1) because of differences in the two sets of optical
constants (Be97 and Ko95) at 7.8 $\mu$m. Compared with model 2, model 3 shows
the same $\varepsilon_2 (\lambda)$ but different $\varepsilon_1 (\lambda)$.
This model was used by Suh \& Kwon (2011) and Jones et al. (2014). Because the
interpolated optical constants are from different materials and do not satisfy
the Kramers-Kronig relation, model 3 would not be physically reliable.

Table 1 summarizes properties of the three models. Even though the three models
can reproduce the observations in similar ways, it is meaningless to use
physically unreliable models for amorphous alumina because other dust species
could also produce similar features in the wavelength range. Therefore, we use
only model 1 for amorphous alumina dust in this work.

\begin{table}
\centering \caption{Models for amorphous alumina dust\label{tbl1}}
\begin{tabular}{llll}
\toprule
Model & Laboratory Data$^{1}$ & K-K$^{2}$	&	Usage$^{1}$	\\
\midrule
model 1 & Be97 & Yes	  & This work	\\
model 2 & Be97; Ko95 & Yes	 & - 	\\
model 3 & Be97; Ko95 & No	  & SK11; Jon14 	\\
\bottomrule
\end{tabular}
\tabnote{$^{1}$Be97: Begemann et al. (1997); Ko95: Koike et al. (1995);
SK11: Suh \& Kwon (2011); Jon14: Jones et al (2014), $^{2}$K-K: the Kramers-Kronig relation.}
\end{table}

\section{Radiative transfer model calculations}

We use the radiative transfer code DUSTY developed by Ivezi\'{c} \& Elitzur
(1997) for a spherically symmetric dust shell. We have performed the model
calculations in the wavelength range 0.1 to 3600 $\mu$m. For all models, we
assume the dust density distribution is continuous ($\rho \propto r^{-2}$). The
dust condensation temperature ($T_c$) is assumed to be 400-1000 K. The outer
radius of the dust shell is always taken to be $10^4$ times the inner radius
($R_c$). We choose 10 $\mu$m as the fiducial wavelength that sets the scale of
the dust optical depth ($\tau_{10}$). For the central star, we assume that the
luminosity is $10^4$ $L_{\odot}$ for all models. We assume that the stellar
blackbody temperature is 2500-3000 K.

For dust opacity, we use a simple mixture of silicate and alumina (10$-$40\% by
number) dust grains as well as pure silicates. For silicate dust, we use the
optical constants of warm and cold silicate grains derived by Suh (1999). We
use the warm silicate dust for LMOA stars (models with $\tau_{10}$ $\leq 3$)
and the cold silicate dust for HMOA stars (models with $\tau_{10}$ $> 3$). For
alumina dust, we use the optical constants of amorphous alumina derived from
this work (model 1; see section 2.2) based on the laboratory data from Be97. We
assume that all dust grains are spherical with a uniform radius of 0.1 $\mu$m.

\subsection{Typical model SEDs}

The upper panel of Figure 2 shows the typical model SEDs for LMOA stars. The
lower panel of Figure 2 shows the typical model SEDs for HMOA stars. They show
the theoretical model SEDs using a simple mixture of silicate and alumina dust
as well as pure silicates. We assume that the stellar blackbody temperature is
2500 K for LMOA stars and 2000 K for HMOA stars. The dust condensation
temperature ($T_c$) is assumed to be 1000 K.

\begin{figure*}[!t]
\centering
\includegraphics[width=120mm]{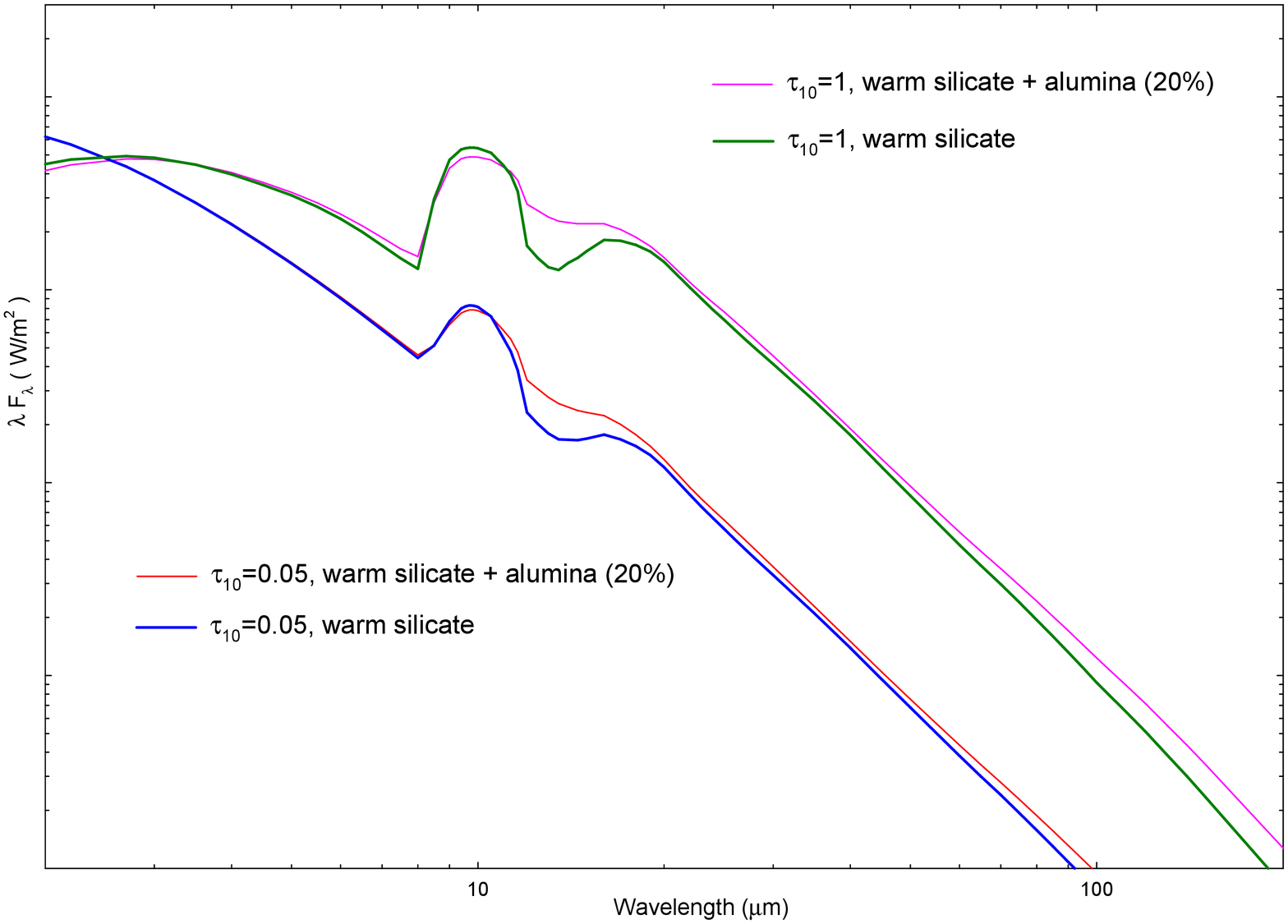} \\
\includegraphics[width=120mm]{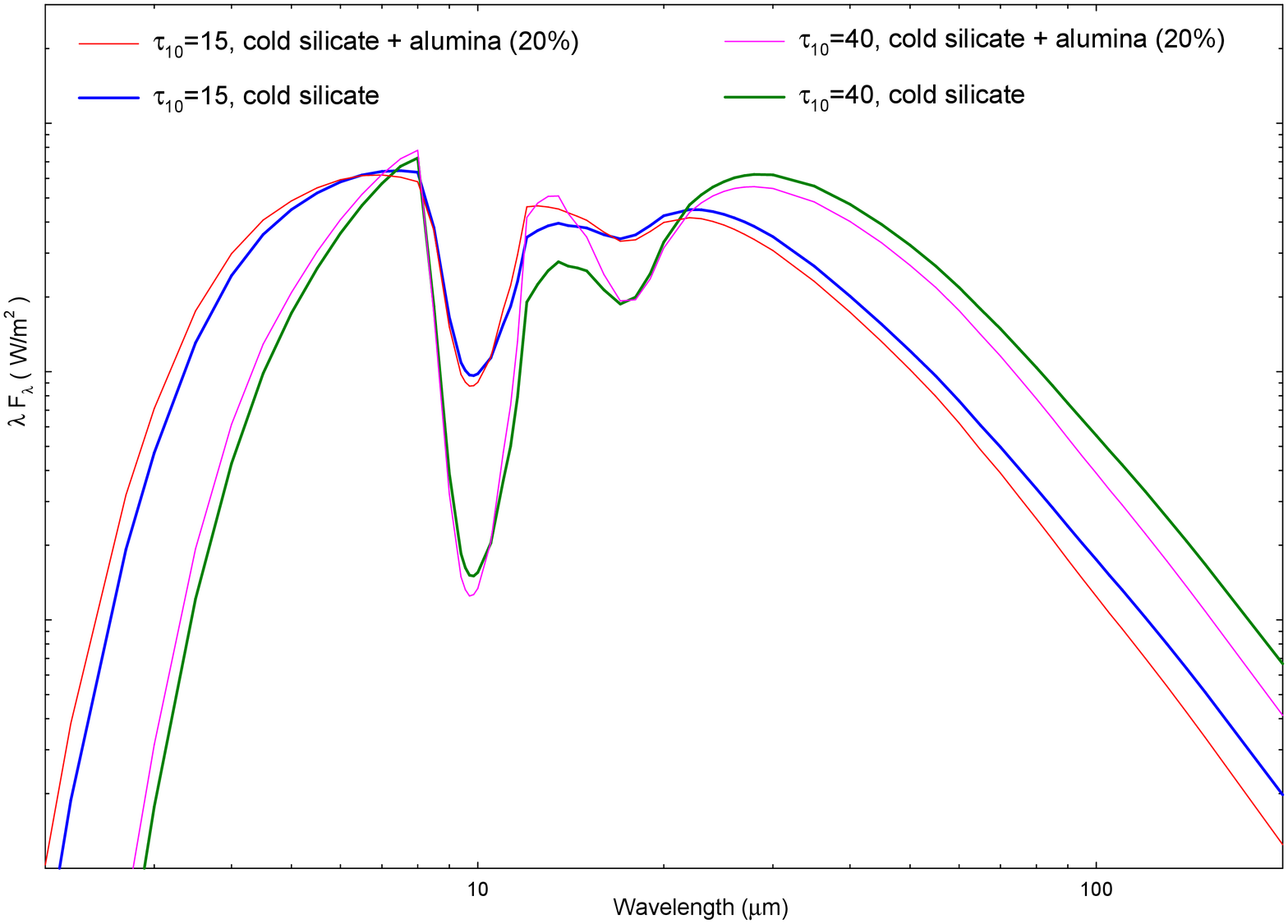}
\caption{Theoretical model SEDs for O-rich AGB stars.}
\end{figure*}

\subsection{Comparison of the model SEDs with the observations}

In Figures 3 and 4, we compare the observed SEDs with the theoretical model
SEDs for the two LMOA stars (omi Cet and Z Cyg) and two HMOA stars (OH
127.8+0.0 and OH 21.5+0.5). For the objects, the spectral data from the
Infrared Space Observatory (ISO) and Infrared Astronomical Satellite (IRAS) Low
Resolution Spectrograph (LRS; $\lambda$ = 8$-$22 $\mu$m) are used. We use the
ISO data from the Short Wavelength Spectrometer (SWS; $\lambda$ = 2.4$-$45.4
$\mu$m) and Long Wavelength Spectrometer (LWS; $\lambda$ = 43$-$197 $\mu$m). We
also use the photometric data from the IRAS Point Source Catalog (PSC) and
AKARI (Murakami et al. 2007).

For LMOA stars, the stellar blackbody temperature is assumed to be 2500 K and
2650 for omi Cet and Z Cyg, respectively. The dust condensation temperature
($T_c$) is assumed to be 654 K and 463 K for omi Cet and Z Cyg, respectively.
Suh (2004) pointed out that $T_c$ looks to be low (400$-$800 K) for LMOA stars
with thin dust envelopes.

For both LMOA stars, a mixture of amorphous alumina with silicate dust grains
produces different model SEDs compared to the model SEDs produced by pure
silicate dust in $\lambda$ = 8$-$20 $\mu$m. The mixture of amorphous alumina
dust looks to reproduces the observed SEDs better in $\lambda$ = 11$-$15
$\mu$m. Other important dust species such as MgFeO series may need to be
considered to make a better fit. The dust grains of MgFeO series produce single
peaks in $\lambda$ = 15-22 $\mu$m (Henning et al. 1995).

In Figure 4, the observed SEDs are compared with the theoretical model SEDs for
HMOA stars. For the two HMOA stars, we find that the amorphous alumina dust is
not useful to reproduce the observed SEDs. For both objects, the stellar
blackbody temperature is assumed to be 2000 K and $T_c$ is assumed to be 1000
K.

\begin{figure*}[!t]
\centering
\includegraphics[width=120mm]{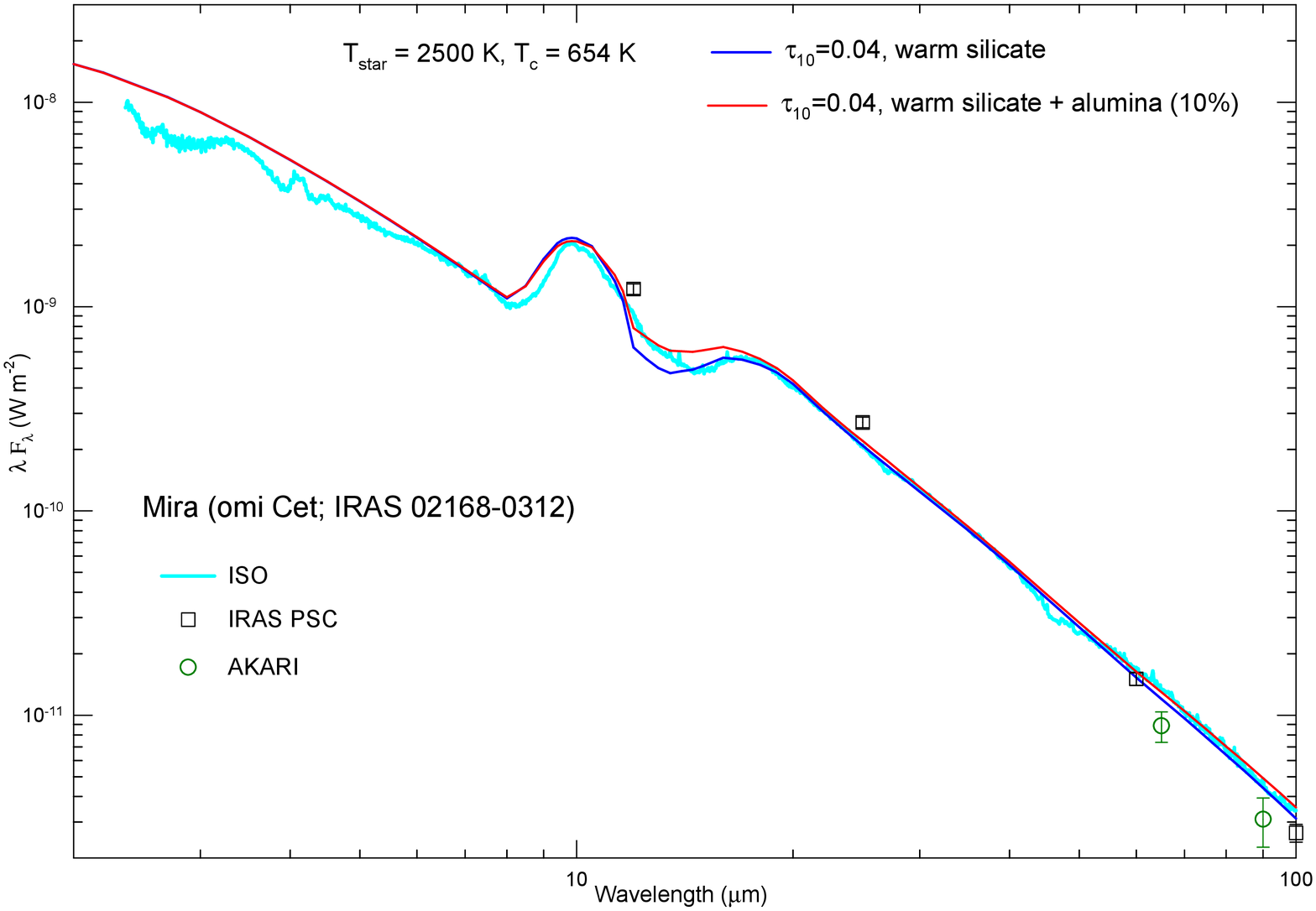} \\
\includegraphics[width=120mm]{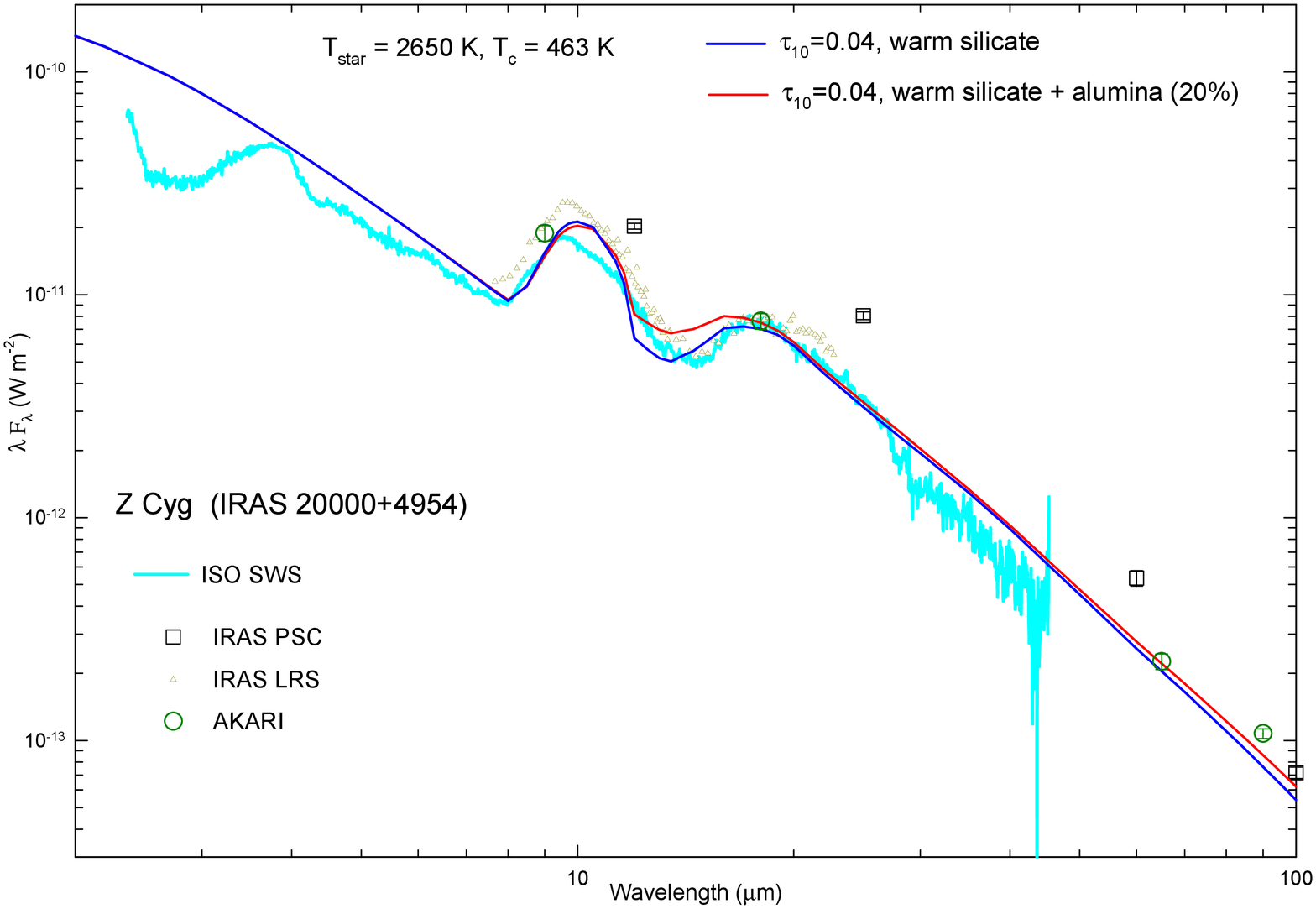}
\caption{Model SEDs compared with the observed SEDs for LMOA stars.}
\end{figure*}

\begin{figure*}[!t]
\centering
\includegraphics[width=120mm]{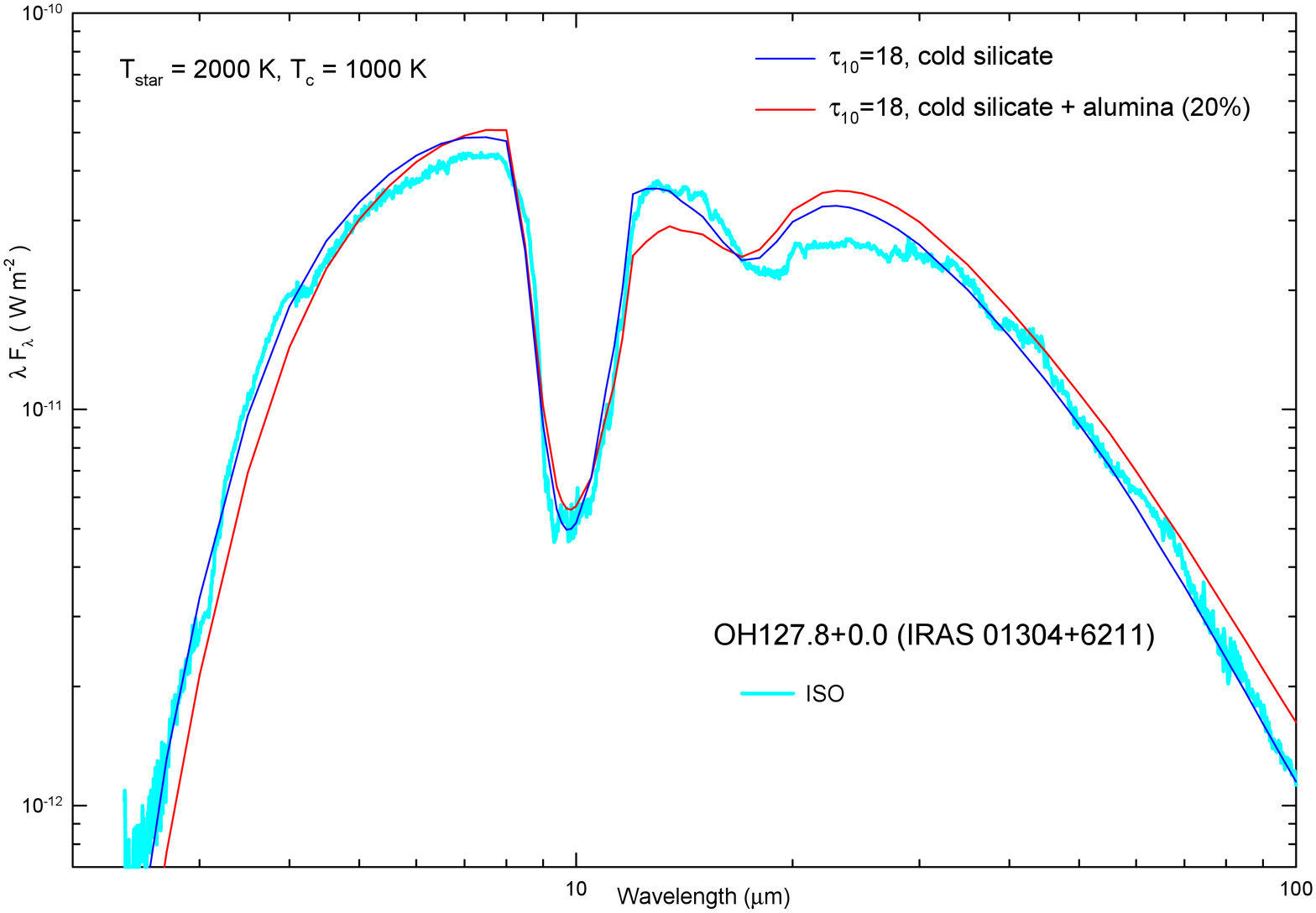} \\
\includegraphics[width=120mm]{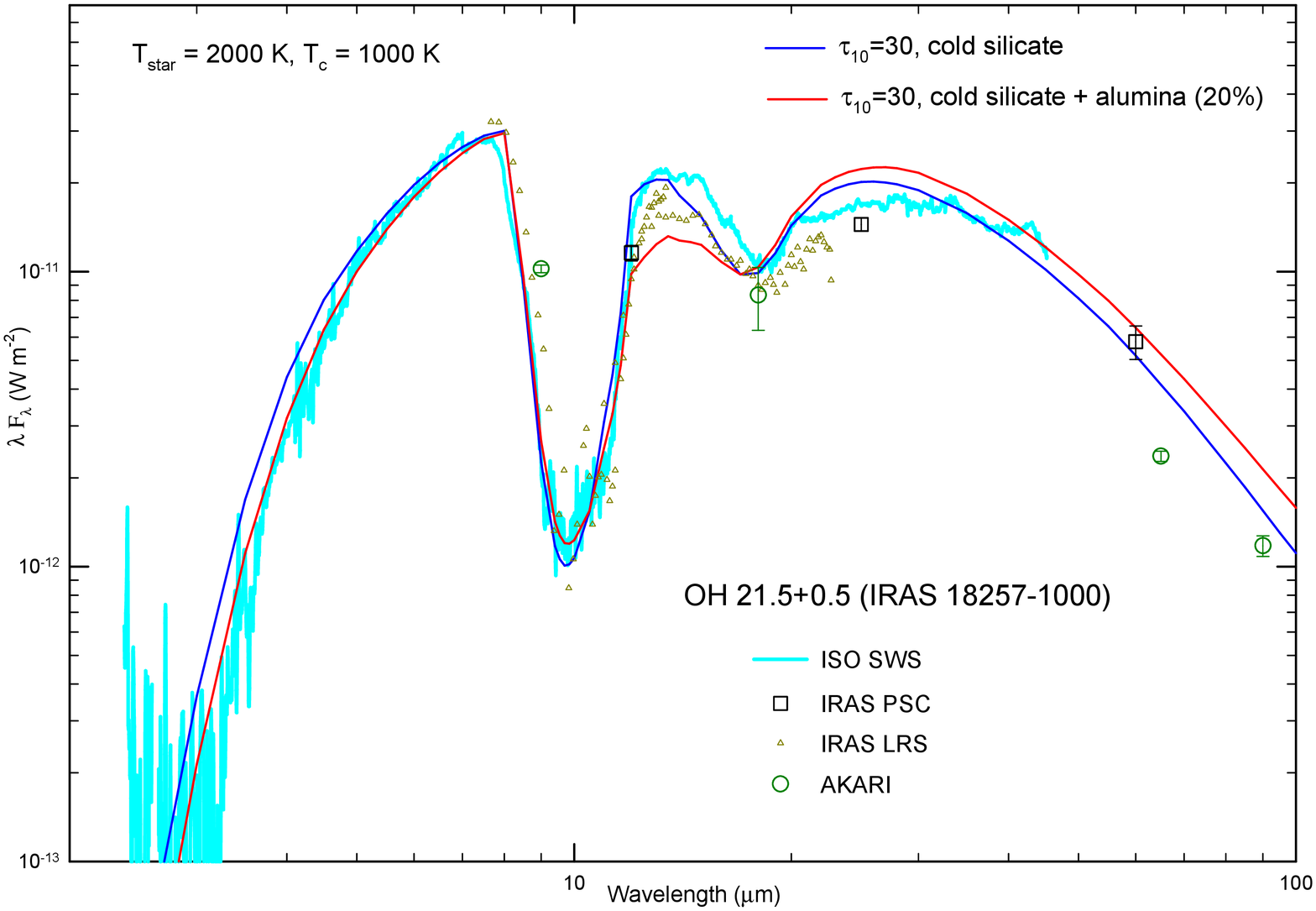}
\caption{Model SEDs compared with the observed SEDs for HMOA stars.}
\end{figure*}

\begin{figure*}[!t]
\centering
\includegraphics[width=140mm]{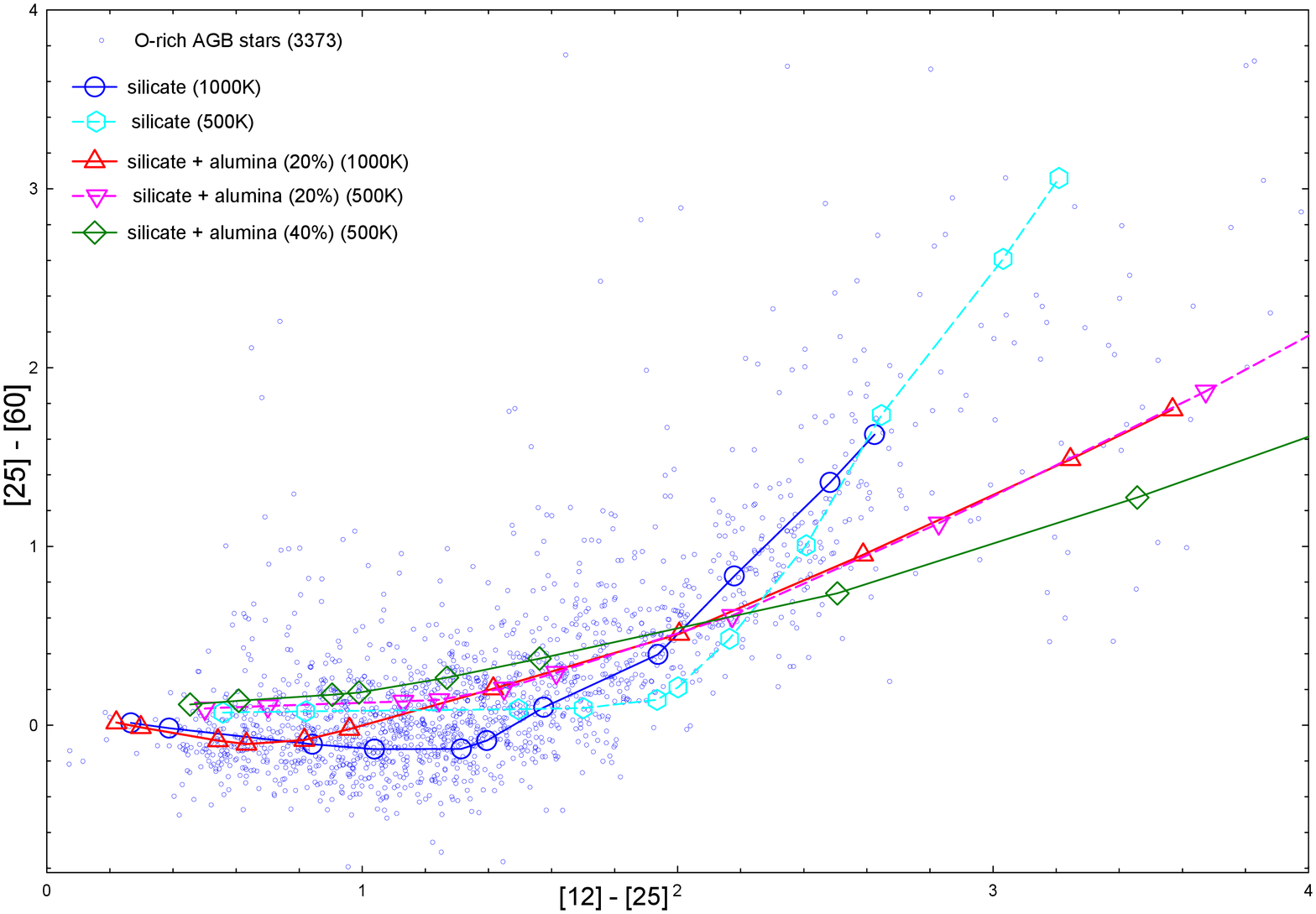} \\
\includegraphics[width=140mm]{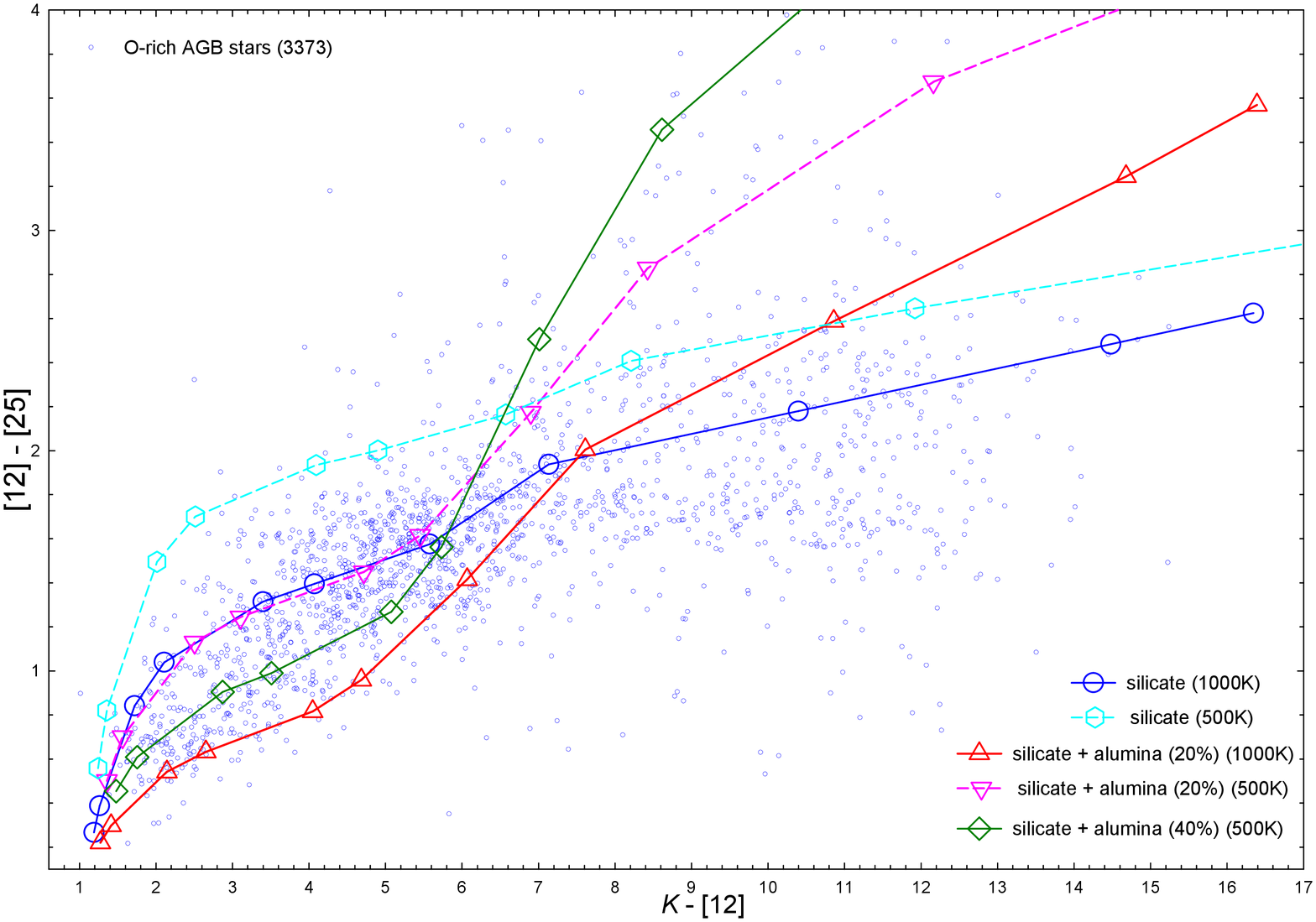}
\caption{IR 2CDs for O-rich AGB stars compared with theoretical dust shell models
(from left to right: $\tau_{10}$ $=$ 0.005, 0.01, 0.05, 0.1, 0.5, 1, 3, 7, 15, 30, and 40).}
\end{figure*}

\begin{figure*}[!t]
\centering
\includegraphics[width=140mm]{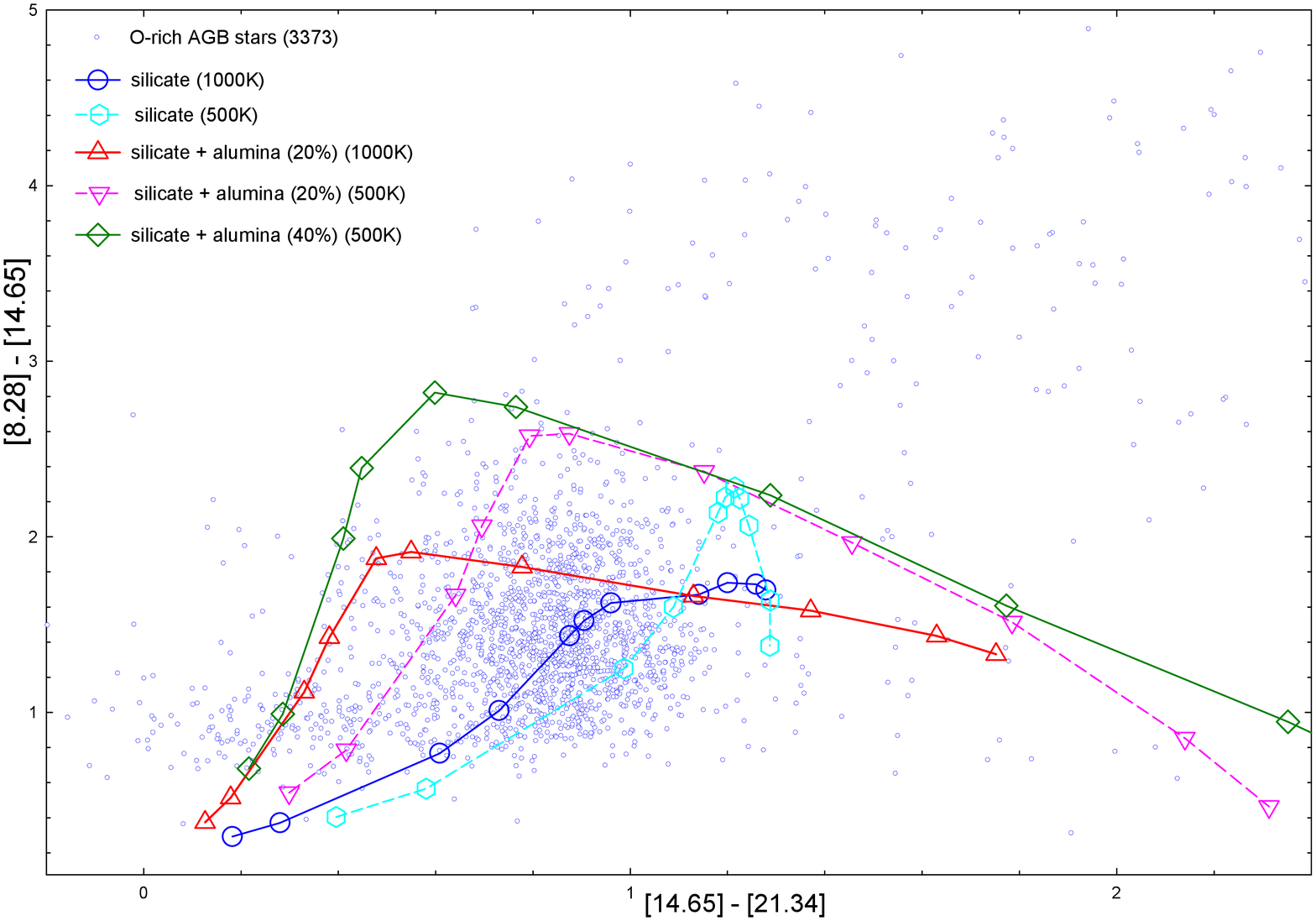} \\
\includegraphics[width=140mm]{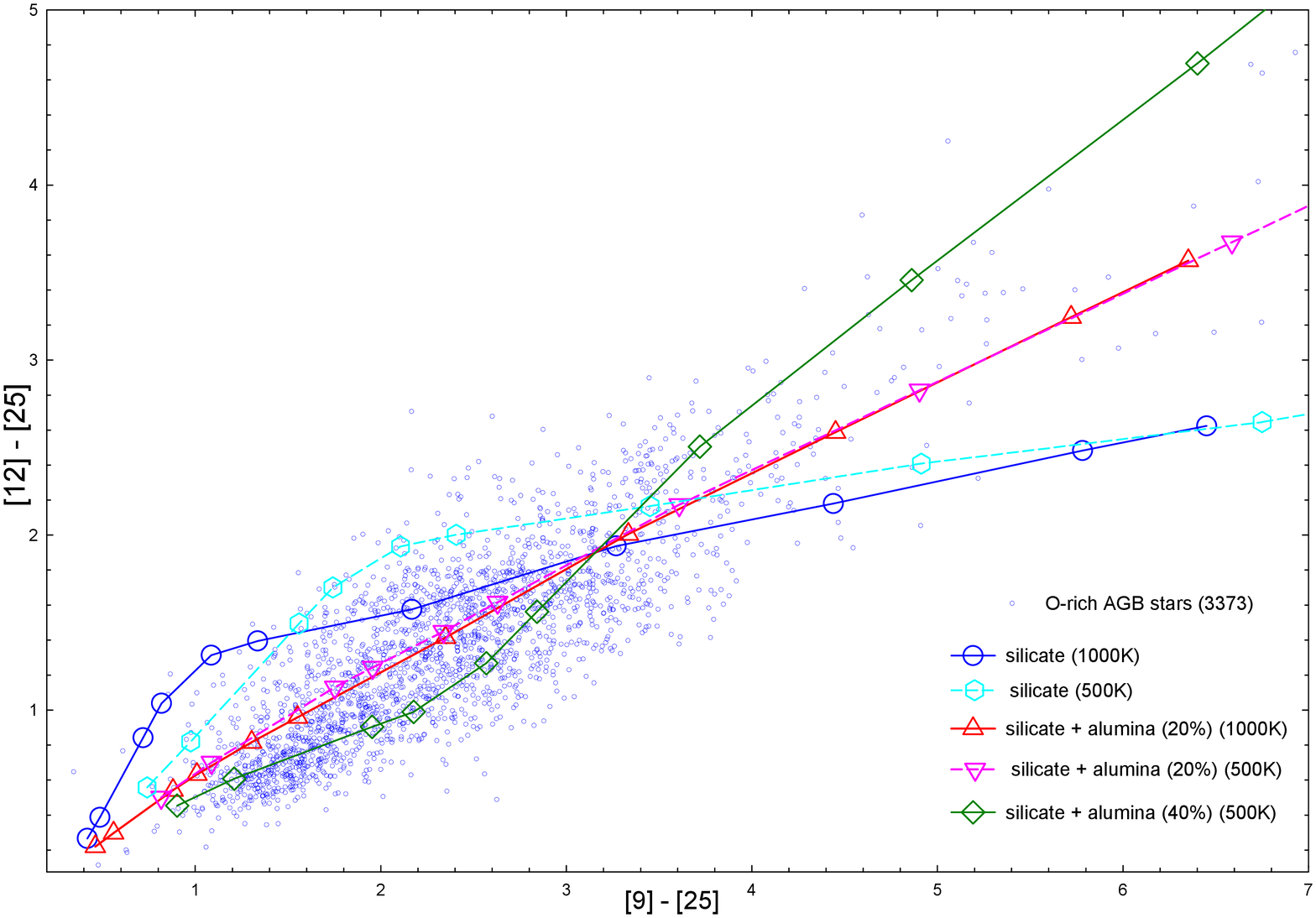}
\caption{IR 2CDs for O-rich AGB stars compared with theoretical dust shell models
(from left to right: $\tau_{10}$ $=$ 0.005, 0.01, 0.05, 0.1, 0.5, 1, 3, 7, 15, 30, and 40).}
\end{figure*}

\section{IR 2CDs for O-rich AGB Stars}

IR 2CDs are useful to characterize the dust envelopes around AGB stars and
post-AGB stars (e.g., Suh 2015). Suh \& Kwon (2011) presented a list of AGB
stars for 3003 O-richstars in our Galaxy and Kwon \& Suh (2012) presented a
revised sample of 3373 O-rich AGB stars.

For the sample O-rich AGB stars, we obtain photometric data in three bands (12,
25, and 60 $\mu$m) by using version 2.1 of the IRAS PSC. We also use the AKARI
PSC data in two bands (9 and 18 $\mu$m) obtained by the infrared camera, Two
Micron All Sky Survey (2MASS; Cutri et al. 2003) data in $K_s$ (2.159 $\mu$m)
band, and Midcourse Space Experiment (MSX; Egan et al. 2003) data in four broad
bands centered at 8.28, 12.13, 14.65, and 21.34 $\mu$m. We cross-identify the
AKARI, 2MASS, and MSX sources by finding the nearest one from the IRAS PSC
position.

The color index is defined by
\begin{equation}
  M_{\lambda 1} - M_{\lambda 2} = 2.5 \log_{10} {{F_{\lambda 2} / ZMC_{\lambda 2}} \over {F_{\lambda 1} / ZMC_{\lambda 1}}}
\end{equation}
where $ZMC_{\lambda i}$ means the zero magnitude calibration at given
wavelength ($\lambda i$) (see Suh \& Kwon 2011 for details). We use only those
objects with good quality data at any wavelength.

Figures 5 and 6 show the IR 2CDs for the 3373 O-rich AGB stars compared with
theoretical models. The small symbols are observational data and the lines with
large symbols are theoretical model calculations for a range in dust-shell
optical depth. We explain the theoretical models in Section 4.1. Generally, the
stars in the upper-right region on the 2CDs have thick dust shells with large
optical depths.

\subsection{Theoretical Models for the 2CDs}

We use the radiative transfer model for the dust shells around O-rich AGB stars
as explained in section 3. We perform the model calculations for eleven optical
depths ($\tau_{10}$ $=$ 0.005, 0.01, 0.05, 0.1, 0.5, 1, 3, 7, 15, 30, and 40).
We assume that the stellar blackbody temperature is 2500 K for $\tau_{10}$
$\leq 3$ and 2000 K for $\tau_{10}$ $> 3$. The dust condensation temperature
($T_c$) is assumed to be 500 K and 1000 K.

We use the warm silicate dust for LMOA stars (7 models with $\tau_{10}$ $\leq
3$) and cold silicate dust for HMOA stars (4 models with $\tau_{10}$ $> 3$). We
also show the models using a simple mixture of silicate and alumina (20$-$40\%)
dust. Generally, the theoretical model points in the upper-right region on a
2CD have large dust optical depths (or thick dust envelopes).

\subsection{Comparison on 2CDs}

Figures 5 and 6 show the IRAS, IRAS-NIR, MSX, IRAS-AKARI 2CDs for O-rich AGB
stars compared with the theoretical models. LMOA stars are located in the
lower-left region and HMOA stars are located in the upper-right region
(lower-right region for the MSX 2CD) in all 2CDs. Generally, we find that the
basic theoretical model tracks roughly coincide with the densely populated
observed points.

The theoretical model tracks are shown for dust condensation temperatures
($T_c$) of 500 K and 1000 K. Suh (2004) pointed out that a lower $T_c$ ($T_c$
$<$ 1000 K) is required for LMOA stars with thin dust envelopes. Generally, we
find that the models with a higher $T_c$ (1000 K) reproduce the observations of
HMOA stars better while the models with a lower $T_c$ (500 K) reproduce the
observations of LMOA stars better for most 2CDs.

The shape of the 10 $\mu$m feature of O-rich AGB stars, which is mainly
produced by silicate, can be modified by addition of the alumina dust. Because
amorphous alumina dust produces a single peak at 11.8 $\mu$m, the AKARI, IRAS,
and MSX fluxes at 9, 8.28, 12, and 14.65 $\mu$m would be easily affected by the
alumina dust. The dust grains of MgFeO series (Henning et al. 1995) produce
single peaks in $\lambda$ 15-22 $\mu$m which could be responsible for the $MSX$
flux at 14.65 and 21.34 $\mu$m. The models with pure silicate dust opacity do
not make satisfactory fits for some 2CDs possibly because of the presence of
alumina and MgFeO series dust.

The upper panel of Figure 5 plots AGB stars in an IRAS 2CD using [25]$-$[60]
versus [12]$-$[25]. The model lines using a mixture of alumina are different
from those using pure silicate because the IRAS flux at 12 $\mu$m is affected
by addition of alumina dust. For HMOA stars, the alumina models are not useful
because they produce exceedingly large [12]$-$[25] colors because the alumina
dust modifies the 10 $\mu$m silicate absorption feature resulting in the
smaller flux at 12 $\mu$m.

The lower panel of Figure 5 shows an IRAS-NIR 2CD using [12]$-$[25] versus
$K$$-$[12]. For LMOA stars, the alumina models are very useful because they
produce smaller [12]$-$[25] colors because the alumina dust modifies the 10
$\mu$m silicate emission feature resulting in the larger flux at 12 $\mu$m. The
alumina models can easily reproduce the LMOA stars in the lower-left region of
the 2CD.

The upper panel of Figure 6 shows a MSX 2CD using [8.28]$-$[14.65] versus
[14.65]$-$[21.34]. The MSX fluxes at 14.65 $\mu$m are affected by the alumina
dust. The alumina models can reproduce the LMOA stars in the upper-left region
because the [8.28]$-$[14.65] colors become redder (or larger) due to the
alumina dust, which produces more emission at 14.65 $\mu$m (see the upper panel
of Figure 2).

On the MSX 2CD, the theoretical model points for very large dust optical depths
are located in lower-right region, unlike other 2CDs. This is because the
[8.28]$-$[14.65] colors become bluer (or smaller) for large dust optical depths
due to the deep silicate absorption feature at 18 $\mu$m, which lower the flux
at 14.65 $\mu$m relative to the flux at 8.28 $\mu$m (see the lower panel of
Figure 2). The objects in the upper-right region could be post-AGB stars (see
Suh 2015).

The lower panel of Figure 6 shows an AKARI-IRAS 2CD using [12]$-$[25] versus
[9]$-$[25]. This 2CD shows the largest deviations of the theoretical models
from the observations because of the AKARI flux at 9 $\mu$m. The similar
effects were noticed in Suh \& Kwon (2011). The observed points in the
lower-left region of the 2CD can be reproduced by the theoretical models using
the larger alumina contents (40-50 \%), compared with other 2CDs.

For all IR 2CDs, we find that the amorphous alumina dust is useful only for
LMOA stars with thin dust envelopes which show the silicate emission features
at 10 and 18 $\mu$m.

\section{Summary} \label{summary}

In this work, we have investigated optical properties of the amorphous alumina
(Al$_2$O$_3$) dust grains in the envelopes around O-rich AGB stars, considering
the laboratory measured optical data. We have derived the the optical constants
of the amorphous alumina dust in a wide wavelength range, which satisfy the
Kramers-Kronig relation and reproduce the laboratory measured data.

Amorphous alumina grains produce a single peak at 11.8 $\mu$m and influences
the shape of the SED at around 10 $\mu$m. The shape of the 10 $\mu$m feature of
O-rich AGB stars, which is mainly produced by silicate, can be modified by
addition of the alumina dust. Using the opacity function of the alumina dust,
we have compared the theoretical radiative transfer model results with the
observed SEDs and observations on various IR 2CDs for a large sample O-rich AGB
stars.

Even though it is difficult to suggest the exact content of amorphous alumina
for O-rich AGB stars because dust species other than alumina can also produce
similar features in the wavelength range 8-15 $\mu$m, we have found a general
trend for a large sample of the stars on 2CDs. Comparing the theoretical models
with the observations on various IR 2CDs, we have found that the amorphous
alumina dust (about 10-40 \%) mixed with amorphous silicate can reproduce much
more observed points for LMOA stars, which have thin dust envelopes. Because
the alumina dust is not useful for HMOA stars, we expect that the relative
alumina abundance for LMOA stars is higher than the abundance for HMOA stars
with thick dust envelopes.

We expect that the optical constants for amorphous alumina derived in this work
would be useful for further studies on dust around AGB and post-AGB stars. The
optical constants for the amorphous alumina derived in this work will be
accessible through the author's web site:
\url{http://web.chungbuk.ac.kr/~kwsuh/d-opt.htm}.

\acknowledgments

This research was supported by Basic Science Research Program through the
National Research Foundation of Korea (NRF) funded by the Ministry of Science,
ICT \& Future Planning (NRF-2013R1A1A2057841). This work was supported by the
intramural research grant of Chungbuk National University in 2015.

\end{document}